\documentclass[usenatbib]{mn2e}
\input{epsf}
\usepackage{color}
\usepackage{graphicx}
\usepackage[utf8]{inputenc}
\definecolor{darkgreen}{rgb}{0,0.7,0}

\renewcommand{\d}[0]{\rm d}

\title
{Why does the Jeans Swindle work?}
\author[M. Falco et al.]{M. Falco$^{1}$,
S. H. Hansen$^{1}$, R. Wojtak$^{1}$ and G. A. Mamon$^{2}$ 
\\
$^1$Dark Cosmology Centre, Niels Bohr Institute, University of Copenhagen,
Juliane Maries Vej 30, 2100 Copenhagen, Denmark;\\
{\rm falco@dark-cosmology.dk},\\
$^2$Institut d'Astrophysique de Paris (UMR 7095: CNRS \& UPMC),
98 bis Bd Arago, 75014 Paris, France}

\begin{document}

\maketitle

\vspace{-1in}

\begin{abstract}
When measuring the mass profile of any given cosmological structure
  through internal kinematics, the distant background density is
  always ignored. This trick is often refereed to as the ``Jeans
  Swindle''. Without this trick a divergent term from the background
  density renders the mass profile undefined, however, this
  trick has no formal justification. We show that when one includes
  the expansion of the Universe in the Jeans equation, a term appears
  which exactly cancels the divergent term from the background. We
  thereby establish a formal justification for using the Jeans
  Swindle.
\end{abstract}


\vspace{-0.45in}

\section{Introduction}

A small overdensity in an otherwise infinite homogeneous gravitating
system (like any cosmological structure in the Universe) is affected
by a basic inconsistency, namely that such a system cannot be in 
equilibrium, and at the same
time obey the Poisson's equation which relates
the gravitational potential to the density distribution. A constant
gravitational potential leads, via Poisson's equation, to a zero density
\citep{J29,ZN71}. The usual
way to overcome this inconsistency is to assume that the infinite
homogeneous system does not contribute to the gravitational potential,
meaning that the gravitational potential is sourced only by
fluctuations to this uniform background density.  This assumption is
called \emph{Jeans Swindle} \citep{bin,bin2,k2003,J2008,E2011}. Following \cite{bin}
\emph{``it is a swindle because in general there is no formal
  justification for discarding the unperturbed gravitational
  field''}. It is vindicated by the right results it provides, but
it is generally considered a limitation to the formalism.

The Jeans Swindle has several applications.  Here we focus on the
Jeans Swindle in the context of the Jeans analysis of internal kinematics, which for instance
is relevant for stellar motions in dwarf galaxies and galaxy motions in
galaxy clusters.  The aim of this work is to explain the ``swindle''
through a clean derivation of the Jeans equation, including
the crucial expansion of the Universe.

The Jeans equation describes systems in equilibrium, and it is therefore
used to model for example dark matter halos inside the virial region,
where they can be treated as equilibrated systems.
Dark matter (DM) halos can be seen as a matter excesses over the mean
matter density of the Universe. This constant background density is
the main contribution to the density distribution at large distances
from the halo center \citep{Tavio08}.  We show that the contributions from the
background density, the cosmological constant and the Hubble
expansion, cancel each other. When omitting the constant background
density (the normal ``swindle'') one is actually excluding it together
with the contribution from the expansion of the Universe. Thus, once we
take into account the expansion of the Universe and the presence of
the cosmological constant, we no longer need to invoke the Jeans
Swindle.

\section{Jeans Swindle in the Jeans equation}

The dynamics of DM halos, modelled as spherical and stationary systems of collisionless particles in
equilibrium, is controlled by the spherical non-streaming Jeans equation \citep{Binney80}
\begin{equation}         
\label{eqn:jeansstd}
-\rho(r)\frac{\d \Phi}{\d r} = \frac{\d (\rho\sigma_{\rm r}^2)}{\d \,r}+2\,\frac{\beta}{r}\rho\sigma_{\rm r}^2 \, ,
\end{equation}
where $\sigma_{\rm r}$ is the radial velocity dispersion, $\beta =
1-\sigma_\theta^2/\sigma_{\rm r}^2$ the
velocity anisotropy, $\rho$ the density distribution of particles
and $\Phi$ the total gravitational potential.

The potential gradient is given by Poisson's equation
\begin{equation}         
\label{eqn:potgrad}
\frac{\d \Phi}{\d r} =\frac{G M(r)}{r^2} \,.
\end{equation}
In the simple case of an isotropic velocity distribution
($\beta=0$), the solution to the standard Jeans equation~(\ref{eqn:jeansstd})
for the radial velocity dispersion is
(from \citealp{Binney80})
\begin{equation}         
\label{eqn:dispersion}
\sigma_{\rm r}^2(r) =\frac{1}{\rho(r)}  
\int_r^\infty\rho(s) 
\,\left[ \frac{G M(s)}{s^2}\right]\d s\, .
\end{equation}
Thus, the only quantity required for the calculation of the radial
dispersion is the density distribution of the halo.  
DM-only cosmological N-body simulations indicate that a double slope profile provides a
reasonable fit to the density profiles of halos within the virial radius
\citep{NFW96,Kra1998}.  Since the integration in equation (\ref{eqn:dispersion}) extends
all the way to infinity, we need a correct description of the DM
distribution beyond the virial radius. The correct asymptotic
value should be the mean matter density of the Universe $\rho_{\rm bg}$,
given by
\begin{equation}
\label{eqn:bgd}
\rho_{\rm bg}=\Omega_M\rho_{c}=\frac{3\,\Omega_M\,H^2}{8\pi\,G} ,
\end{equation}
where $\Omega_M$ the matter density parameter, $\rho_c$ is the
critical density of the Universe and $H=\dot{a}/a$ is the Hubble constant ($a$ being the scale factor
of the Universe). 

\begin{figure}
\begin{center}
    \leavevmode
    \epsfxsize=6.5cm
    \epsfbox[100 30 480 400]{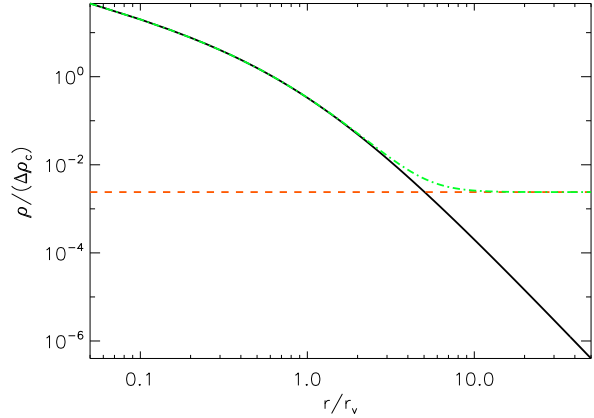}
\end{center}
	\caption{\emph{Black solid line}: Hernquist profile given by
          equation~(\ref{eqn:rhohern}). \emph{Red dashed line}: background
          density given by equation~(\ref{eqn:bgd}). \emph{Green
            dash-dot line}: the sum of the
        Hernquist profile and the constant background density.
}
\label{fig:density}
\vspace{-0.2in}
\end{figure}

Therefore, the double slope profile, reaching zero density at large distances
from the cluster center, does not reproduce the right density profile in the
external region \citep{Tavio08}.  As a first approximation, we can write the density
as given by the sum of a term $\rho_{\rm h}$ that reproduces the inner part
of the halo distribution and the constant background density that
affects the profile only at large radii
\begin{equation}
\label{eqn:rhotot}
\rho(r)=\rho_{\rm h}(r)+\rho_{\rm bg} \, .
\end{equation}
As an example, we consider a finite mass density 
profile for a cluster-size halo, the \citet{Hern90}
profile~\footnote{Here we discuss the simple Hernquist profile for academic reasons. Using
  any other finite mass structure would lead to the same conclusions.} 

\begin{equation}
\label{eqn:rhohern}
\rho_{\rm h}(r)=\frac{\rho_0}{r/r_{\rm v}\,(1+r/r_{\rm v})^3}\, ,
\end{equation}
where $r_{\rm v}$ is the virial radius and $\rho_0$ is the
characteristic density, that can be written in terms of the virial
overdensity $\Delta$ as

\begin{equation}
\label{eqn:rhochar}
\rho_0=\alpha\,\Delta\rho_c\,.
\end{equation}
In Figure~\ref{fig:density}, we show the profile given by eq.~(\ref{eqn:rhohern}) (black solid line)
and the sum $\rho_{\rm h}(r)+\rho_{\rm bg}$ (green dash-dot line), where the asymptotic
value is $\rho_{\rm bg}$ (red dashed line).  In
the calculation, we
set $\Omega_M=0.24$, $\Delta=100$, $H=73\,\rm km\,s^{-1}\,Mpc^{-1}$
and we fix $\alpha$ by imposing that the mass given by the density in
eq.~(\ref{eqn:rhohern})  within the sphere of radius $r_{\rm v}$
corresponds to the virial mass 

\begin{equation}
\label{eqn:virialmass}
M_{\rm v}=\frac{4\pi}{3}r_{\rm v}^{3}\Delta\rho_{c}\, .
\end{equation} 
In Figure~\ref{fig:density}, the density is in units of $\Delta\rho_c$
and the radius is in units of the virial radius.

\begin{figure}
\begin{center}
    \leavevmode
    \epsfxsize=6.5cm
    \epsfbox[100 30 480 400]{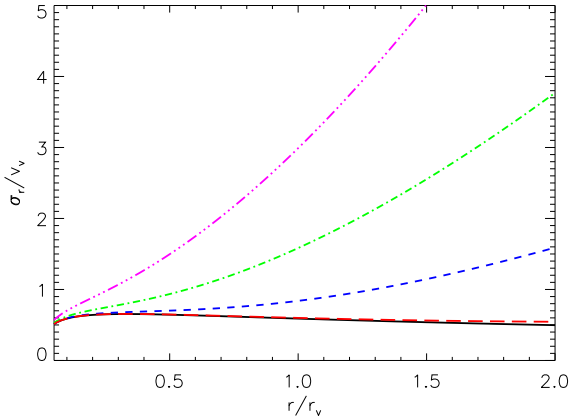}
\end{center}
	\caption{Radial velocity dispersion profiles. The black solid line
          corresponds 
to the standard Jeans solution~(\ref{eqn:sig0}). Solutions of
          equation~(\ref{eqn:sig}) corresponding to different upper
          limits $r_{\rm max}$ in the integral are shown: $r_{\rm max}=200\,r_v$ (blue short-dashed
          line), $r_{\rm max}=500\,r_v$ (green dash-dot line), $r_{\rm max}=1000\,r_v$
          (magenta dash-dot-dot
          line). The red long-dashed line is the solution when including
          cosmological expansion and background density
          (eq.~[\ref{eqn:dispersiontot}]). The radial velocity
          dispersion is in units of the virial velocity and the radius is
          in units of the virial radius.}
\label{fig:sigma}
\vspace{-0.2in}
\end{figure}

Once we insert the expression (\ref{eqn:rhotot}) in the integral
(\ref{eqn:dispersion}) we get 

\begin{eqnarray}
\label{eqn:sig}
\sigma_{\rm r}^2(r) &\!\!=\!\!&\frac{1}{\rho_{\rm h}(r)\!+\!\rho_{\rm bg}} 
 \nonumber \\ &\!\!&\times 
\,
\int_r^\infty \!\!G \,\left[\rho_{\rm h}(s)\!+\!\rho_{\rm bg}\right] 
\left[ M_{\rm h}(s)\!+\!M_{\rm bg}(s) \right]
\frac{\d s}{s^2} \,,
\end{eqnarray}
which diverges, since the background mass diverges at large radii.
In Fig.~\ref{fig:sigma} we plot the solution~(\ref{eqn:sig}) for different upper
limits $r_{\rm max}$ in the integral: $r_{\rm max}=200\,r_{\rm v}$ (blue short-dashed line),
$r_{\rm max}=500\,r_{\rm v}$ (green dash-dot
line), $r_{\rm max}=1000\,r_{\rm v}$ (magenta dash-dot-dot line). 
 
This clearly indicates that when
integrating to infinity, the integral will diverge.

The usual trick to avoid the divergence is to 
omit the contribution of
the background density to the potential gradient
(\ref{eqn:potgrad}), i.e. to set 
$\nabla\Phi_{\rm bg}=0$ 
\citep{bin,bin2,E2011}. Physically, this amounts
to assume that  the
gravitational potential is sourced only by fluctuations to the
uniform background density. For this
requirement to be consistent with the Poisson's equation
(\ref{eqn:potgrad}), the constant
$\rho_{\rm bg}$ in equation (\ref{eqn:rhotot}) has to vanish. 
 This assumption is called the \emph{Jeans Swindle}. It has no
 justification other than to overcome a mathematical difficulty.

 The Jeans analysis has always been performed by discarding the
 \emph{unperturbed} density, so that equation (\ref{eqn:dispersion})
 is just
\begin{equation}
\label{eqn:sig0}
\sigma_{\rm r}^2(r)=\frac{1}{\rho_{\rm h}(r)} 
\int_r^\infty\rho_{\rm h}(s) \left[ \frac{G\,M_{\rm h}(s)
  }{s^2}\right]\d s\, .
\end{equation}
The black solid line in fig.~\ref{fig:sigma}  corresponds to the computation of
equation~(\ref{eqn:sig0}), which is not divergent.

With this approach, the Jeans equation reproduces the radial
dispersion of DM halos from cosmological simulations in the region of
equilibrium \citep{SM04,WLGM05,CPKM08}, and gives finite masses for
dwarf galaxies \citep{sfw10,sww12}.

\section{Why the Jeans Swindle works}

We wish to replace the Jeans Swindle by a formally correct analysis.
Thus, we keep the background density $\rho_{\rm bg}$ and its contribution
in the gravitational potential.  The gradient of the potential due to
$\rho_{\rm bg}$ can be put in the following form:
\begin{equation}
\frac{\d \Phi_{\rm bg}}{\d r} =\frac{4}{3}\pi\,G\rho_{\rm bg}\,r={\Omega_{\rm
    m}\over 2}\,H^2r\, ,
\end{equation}
where we have used equation (\ref{eqn:bgd}).  However, in order to be
consistent, we need to take into account all effects due to the
underlying cosmology.
For the case of a single halo embedded in a homogeneous Universe with a non-zero cosmological constant
$\Lambda$, particles also feel a repulsive potential of the form
(e.g. \citealp{PP2006,Nandra12})
\begin{equation}
\frac{\d \Phi_{\rm \Lambda}}{\d r}
=-\frac{1}{3}\Lambda\,r=-\Omega_{\Lambda}\,H^2\,r\, ,
\end{equation}
where we have used the relation
\begin{equation}
\Omega_{\Lambda}=\frac{\Lambda}{3\,H^2}\, .
\end{equation}
Introducing the deceleration parameter
\begin{equation}
\label{eqn:q}
 q=-{\ddot{a}a \over \dot{a}^2}={\Omega_{\rm m}\over 2}-\Omega_{\Lambda},
\end{equation}
we can rewrite the total contribution of the cosmology to the
gravitational potential gradient as
\begin{equation}
\frac{\d \Phi_{\rm bg}}{\d r}+\frac{\d \Phi_{\Lambda}}{\d r} =q\,H^2\,r.
\end{equation}
Moreover, the Universe is not static, but it is subject to the Hubble
expansion.  Equation (\ref{eqn:jeansstd}) holds for structures
that have achieved dynamical equilibrium. This means that the radial,
longitudinal and azimuthal bulk motions are not taken into account in
its derivation.  When excluding all these bulk velocity terms,
the Hubble flow, which DM particles are subject to, is also discarded. The
Hubble velocity, $v_H=H\,r$, might be neglected in the very inner
region, but for large radii, it becomes important. Since the
integration in equation (\ref{eqn:dispersion}) extends to infinity,
the inclusion of $v_H$ will affect the result.

When we include the terms involving the mean radial velocity, the
Jeans equation becomes the more general formula \citep{F2012}
\begin{equation}
\label{eqn:jeansgen2}
-\rho\frac{\d \Phi}{\d r}=\frac{\d (\rho\sigma_{\rm r}^2)}{\d 
  r}+2\,\frac{\beta}{r}\rho\sigma_{\rm r}^2 
+\rho \left[\overline v_{\rm r}\frac{\partial \overline v_{\rm r}}{\partial r}+\frac{\partial \overline v_{\rm r}}{\partial t} \right]\, .
\end{equation}
In the most general case, $v_{\rm r}$ is the sum of the Hubble velocity and
a peculiar infall velocity. 
The infall velocity occurs around
cluster-sized haloes ($M_{\rm vir}\approx\,10^{13-14}M_{\odot}$),
and is totally negligible around galactic haloes ($M_{\rm
  vir}\approx\,10^{12}M_{\odot}$) \citep{Prada06,CPKM08}.
Streaming motions around clusters are dominated by the infall velocity
at radii between the virial radius and the turn-around radius, which
is approximately equal to $3.6$ virial radii  \citep{CMM08}. At larger distances,
it approaches the Hubble flow. Therefore, far outside the equilibrated cluster, we
can neglect the mean radial peculiar motion of particles, so that the radial velocity corresponds to
$v_H$ only
\begin{equation}
\label{eqn:vel}
\overline v_{\rm r} (r,t)=H(t)\,r\, .
\end{equation}
It is straightforward to calculate the additional term in square
brackets in eq.~(\ref{eqn:jeansgen2})
\begin{equation}
\label{eqn:pecvel}
\overline v_{\rm r}\frac{\partial \overline v_{\rm r}}{\partial r}+\frac{\partial
  \overline v_{\rm r}}{\partial t} = H^2\,r+\dot{H}\, r=-q\,H^2r \, ,
\end{equation}
where we used
\begin{equation}
\label{eqn:hdot}
\dot{H}=-(q+1)\,H^2\, .
\end{equation}
We can now write equation (\ref{eqn:dispersion}) {\bf for large radii}, including all these
cosmological terms
\begin{eqnarray}
\label{eqn:dispersiontot}
\sigma_{\rm r}^2(r) \!\!\!&\!\!\!=\!\!\!&\!\!\!\frac{1}{\rho_{\rm h}(r)\!+\!\rho_{\rm
    bg}}\int_r^\infty\left [\rho_{\rm h}(s) \!+\!\rho_{\rm bg} \right ]
 \nonumber \\ &&\times 
\,
\left[
  \frac{\d \Phi_{\rm h}}{\d s}\!+\!\frac{\d \Phi_{\rm bg}}{\d s}\!+\!\frac{\d \Phi_{\Lambda}}{\d s}\!+\!\overline v_r\frac{\partial \overline v_r}{\partial s}\!+\!\frac{\partial \overline v_r}{\partial t}\right]\d s \nonumber  \\
\!\!\!&\!\!\!=\!\!\!&\!\!\! \frac{1}{\rho_{\rm h}(r) \!+\!\rho_{\rm bg}} \int_r^\infty
\left [\rho_{\rm  h}(s) \!+\!\rho_{\rm bg} \right ]
\nonumber \\ &&\times 
\,
\left[ \frac{G
 M_{\rm h}(s)}{s^2}\!+\! q\,H^2\,s|_{\rm bg,\Lambda}-q\,H^2\,s|_{H}\right]\d s
\nonumber \\
\!\!\!&\!\!\!=\!\!\!&\!\!\!\frac{1}{\rho_{\rm h}(r) \!+\!\rho_{\rm bg}}
\int_r^\infty(\rho_{\rm h}(s)\!+\!\rho_{\rm bg}) 
\,
\left[ \frac{G M_{\rm h}(s)}{s^2}\right]\d s\, .
\end{eqnarray}
We thus see that the term $-q\,H^2\,s|_H$ given by the Hubble velocity
(\ref{eqn:pecvel}) cancels exactly the term $q\,H^2\,s|_{{\rm bg},\Lambda}$ 
given by the potentials of
the background density and the cosmological constant.  In this way, we
recover the same result as applying the Jeans Swindle, and in the
Jeans solution the total mass is again $M(s)=M_{\rm h}(s)$.

Formally, there is still a minor difference between the two approaches: the
density involved in eq.~(\ref{eqn:dispersiontot}) is still given by
$\rho_{\rm h}(s)+\rho_{\rm bg}$, where $\rho_{\rm bg}$ is not zero
but instead given by eq.~(\ref{eqn:bgd}). However, this time it does not lead to
any divergence, because $M_{\rm h}(s)/s^2$ falls rapidly to zero at large distances. 
This can be seen in
Figure~\ref{fig:sigma}, where the solution of eq.~(\ref{eqn:dispersiontot})
is the red long-dashed line and it matches the result we obtain from
equation~(\ref{eqn:sig0})(black solid line). For larger radii, the
addition of the background density in the density profile can affect
the result slightly. However, in the outer regions, where the halos
are no longer equilibrated, the standard Jeans equation is anyway not used to
reproduce the radial velocity dispersion. Instead, the generalized Jeans
equation in  eq.~(\ref{eqn:jeansgen2}) must be used, including the infall
motion of galaxies. The addition of the peculiar velocity changes the
shape of the velocity dispersion in the infall
region \citep{F2012}, but it does not affect the conclusion of this work.
One could also improve on this minor difference, by not including
$\rho_{\rm bg}$ at all radii, but instead a different form which takes
into account that the immediate environment of haloes may not be the
cosmological value yet. A more accurate density profile would include
a term to describe the local
region around clusters, before the cosmological background is
reached. For example, \cite{CS02} give a detailed description of the halo
model, where the background contribution to the total density is given by a more complicated
function than the constant value $\rho_{\rm bg}$ only.
This is equivalent to define $\rho_{\rm h}$ as 
\begin{equation}
\rho_{\rm h}=\rho_{\rm tot}-\rho_{\rm bg}\, ,
\end{equation}
i.e. including in $\rho_{\rm
  h}$ the details of the background
density being different from $\rho_{\rm bg}$. 
The equation~(\ref{eqn:dispersiontot}) is
formally not affected by this modification.

\section{Comoving coordinates}

The equations describing the particle distribution and
motion can be written in comoving
coordinates \citep{Pee80}. The physical coordinates $r$ and comoving
coordinates $x$ are related by the universal time-dependent expansion parameter $a(t)$ 
\begin{equation}
r=a(t)\,x\, .
\end{equation} 
When changing variables from the physical space to the comoving one,
the Poisson's equation becomes \citep{Pee80}
\begin{equation}
\nabla^2\phi=4\pi\,G\,a^2[\rho(x)-\rho_{\rm bg}]\, , 
\end{equation} 
where the gradient is with respect to $x$, and $\rho_{\rm bg}$ is the mean
mass density and $\phi(x)$ is the potential contributed by the
overdensity $\rho-\rho_{\rm bg}$. Therefore, in this
coordinate system, the particle motion is already described in terms of the departure
from the constant background, and the swindle is not required. As we expect,
taking into account the cosmological expansion in the physical space
leads to the same result as moving to the expanding space. 
Equation~(\ref{eqn:jeansstd}) 
would be correct if we replaced $r$ with
$x$ and $\Phi$ with $\phi$, and using $\rho$ given by~(\ref
{eqn:rhotot}), namely it is the correct Jeans equation in comoving coordinates. 

\cite{J2012} have also shown that a cosmological N-body simulation of
an isolated overdensity should reproduce, in physical coordinates, the same result as
a simulation obtained for the structure in open boundary condition without expansion.

We conclude that our work is consistent with the comoving frame
analysis by \cite{Pee80} and with the conclusions of \cite{J2012} . This confirms that the Jeans Swindle corresponds to accounting for the
expansion of the Universe. 

\vspace{-0.25in}

\section{Conclusions}

We have demonstrated that the Jeans Swindle is not an \emph{ad hoc}
trick, but it is the result of correctly combining the mean matter
density and the expansion of the Universe. The divergent term from the
background density, which in a static universe would lead to a
divergent dispersion profile, is exactly cancelled by a term from the
expanding universe. We have shown that the dispersion profile measured
when assuming no background and a static universe, is the same as the
dispersion profile when including both the background density and the
expansion.  This means that we have establish a formal justification
for using the Jeans Swindle.  This result holds for radii smaller than
roughly the virial radius. For larger radii one has to include the
effect of infalling matter, which is done through a generalized Jeans
equation, as will be presented in a forthcoming article \citep{F2012}.

\vspace{-0.25in}

\section*{Acknowledgements}

We thank Wyn Evans for comments, Michael Joyce for useful discussions, and the referee Mark Wilkinson for
comments which improved the letter.
The Dark Cosmology Centre is funded by the Danish National Research Foundation.

\vspace{-0.25in}

\bibliography{swindle}

\end{document}